# Orthogonal electric control of the out-of-plane field-effect in two-dimensional ferroelectric α-In$_2$Se$_3$

Yue Li[a,b,1], Chen Chen[a,b,1], Wei Li[b], Xiaoyu Mao[a,b], Heng Liu[a,b], Jianyong Xiang[c], Anmin Nie[c], Zhongyuan Liu[c], Wenguang Zhu[a,b,2] & Hualing Zeng[a,b,2]

[a.] International Center for Quantum Design of Functional Materials (ICQD), Hefei National Laboratory for Physical Science at the Microscale, and Synergetic Innovation Center of Quantum Information and Quantum Physics, University of Science and Technology of China, Hefei, Anhui 230026, China

[b.] Key Laboratory of Strongly-Coupled Quantum Matter Physics, Chinese Academy of Sciences, Department of Physics, University of Science and Technology of China, Hefei, Anhui 230026, China

[c.] State Key Laboratory of Metastable Materials Science and Technology, Yanshan University, Qinhuangdao 066004, China

[1] Contribute equally to this work

[2] Corresponding author

Abstract: Tuning the electric properties of crystalline solids is at the heart of material science and electronics. Generating the electric field-effect via an external voltage is a clean, continuous and systematic method. Here, utilizing the unique electric dipole locking in van der Waals (vdW) ferroelectric α-In$_2$Se$_3$, we report a new approach to establish the electric gating effect, where the electrostatic doping in the out-of-plane direction is induced and controlled by an in-plane voltage. With the vertical vdW heterostructure of ultrathin α-In$_2$Se$_3$ and MoS$_2$, we validate an in-plane voltage gated coplanar field-effect transistor (CP-FET) with distinguished and retentive on/off ratio. Our results demonstrate unprecedented electric control of ferroelectricity, which paves the way for integrating two-dimensional (2D) ferroelectric into novel nanoelectronic devices with broad applications.

# 1. Introduction

Field effect, the ability of electrically modulating carrier density in solid material, is not only the fundamental of semiconductor based information technology but also crucial for studying a wide variety of field-induced rich phenomena in fundamental science, such as unconventional superconductivity,[1-3] exotic magnetism,[4] emergent phase transition,[5-7] and topological quantum transport.[8, 9] To generate electric gating, the universal scheme is constructing the device referred to as field-effect transistor (FET), in which an external voltage can be applied to vary the carrier doping of the material in the conducting channel. To date, there have been delivered several types of FETs.[5, 7, 10-14] Subject to specific gate dielectric, the FETs function in distinctive way, including electrostatically accumulation of carriers via oxide insulator or ferroelectrics,[12-14] forming the electric double layer of large capacitance by organic ionic liquids,[10, 11, 15] and driving ion intercalation from recently developed solid ionic gel or conductor.[5, 7] However, limited by the linear electric response of conventional gate dielectrics, the device geometry in most field-effect studies is fixed to be capacitor-like by placing an indispensable gate electrode either on top or at bottom to sandwich the dielectrics. The field-effect is therefore induced by the gate electric field, which is always perpendicular to the conducting channel. This stereo structure geometry consequently increases the device complexity and furtherly limits its application in specific areas only. To that end, FETs with new conceptual design is highly desired, especially for developing flexible nanoelectronics and optoelectronics with van der Waals (vdW) materials.

In this work, to overcome the limitation, we present a new approach to create a vertical field-effect in a co-planar device with layered α-In$_2$Se$_3$ as the gate dielectric. α-In$_2$Se$_3$ is an emerging two-dimensional (2D) ferroelectric semiconductor with remarkable optical and transport properties.[16-19] Recent studies have confirmed its stable room-temperature ferroelectricity at the single atomic layer limit with the unique intercorrelation of out-of-plane and in-plane electric polarizations.[20-26] For ferroelectrics, especially when forming heterostructures, their polarity is an effective knob to tune the electrostatic doping as well as the global electronic structure, resulting practical device potential such as ferroelectric field-effect transistors (FeFETs).[27] This same scenario is applicable to all 2D ferroelectrics with OOP electric polarizations.[13, 28, 29] In additional to the OOP ferroelectricity, α-In$_2$Se$_3$ presents the unique interlocking feature of electric dipoles. The flipping of the IP electric polarization in α-In$_2$Se$_3$ can be achieved by an OOP bias,[21, 24] or vice versa. Therefore, it suggests that one can develop co-planar FET (CP-FET) with ultrathin α-In$_2$Se$_3$, where the out-of-plane (OOP) electric gating effect is controlled by an in-plane (IP) voltage.

Our results are summarized in the following. We verify that the IP and OOP electric polarization in 3R α-In$_2$Se$_3$ thin layers are inherently coupled, which is consistent with previous studies.[20, 21, 24] Such correlation is observed in both spontaneous and artificial ferroelectric domains. Based on the electric dipole locking, we experimentally show the reversible switch of the OOP electric polarization via the bias in a two-electrode planar device of layered α-In$_2$Se$_3$ only. In particular, we find large surface potential difference (~100 mV) for oppositely polarized ferroelectric domains. This can be well understood as a result of the accumulated negative or positive polarization charges on the domain surface. We fabricate the CP-FET by stacking the vertical vdW heterostructure with ultrathin MoS$_2$ and α-In$_2$Se$_3$. In the device, the conductance of few layer MoS$_2$ can be reversibly tuned by the in-plane polarizing voltage applied on α-In$_2$Se$_3$. With the non-volatile nature of ferroelectrics, the as-prepared CP-FET functions with distinguished and retentive on/off ratio. The findings in this work demonstrate a new way of manipulating ferroelectricity, which enables the unprecedented function control with simple device architecture.

## 2. Results and Discussion

Figure 1 depicts the schematic lattice structure of α-In$_2$Se$_3$ and the corresponding ferroelectricity. Bulk α-In$_2$Se$_3$ holds rhombohedral *R3m* layered crystal structure (Figure 1a). Inside the unit cell, there are three quintuple layers (QLs), which contains five covalently bonded atomic layers in the sequence of Se-In-Se-In-Se. In each α-In$_2$Se$_3$ QL, the central Se atoms are asymmetrically bonded to four In atoms, forming tetrahedron structure. Consequently, the interlayer spacings between the middle Se layer and its neighboring two In layers are dramatically different. This subtle atomic configuration of the central Se atoms leads to the emergence of ferroelectricity in α-$In_2Se_3$. As the shown in Figure 1b, by projecting all the atoms of a given QL onto c axis and a axis ( or b axis), it is clear to find that the spatial inversion symmetry is explicitly broken along both IP and OOP directions. The deviation of the middle Se atom from the inversion center contributes to the generation of net electric dipole in the OOP direction as well as in the IP direction. With 180 degree in plane rotation of the QL, while remianing the lattice structure and symmetry unchanged, the IP and OOP electric polarizations are reversed, establishing the degenerate ferroelectric states in each direction. Most importantly, the crystal lattice of α-$In_2Se_3$ guarantees the intercorrelation between the IP and OOP electric polarizations, which makes it possible to manipulate the electric polarization in one direction by controlling the other. This is the unique feature of 2D α-$In_2Se_3$ that is absent in intensively studied ferroelectrics, such as perovskites oxides $BaTiO_3$, the organic ploy(vinylidene fluoride-co-trifluoroethylene) (PVDF-TrFE) thin film and recently discovered 2D $CuInP_2S_6$.

The α-$In_2Se_3$ flakes studied here were mechanically exfoliated from bulk crystals (details can be found in Methods). To confirm its crystalline structure, we characterized our samples with aberration-corrected scanning transmission electron microscopy (AC-STEM) and the Raman technique. Figure 1c shows a typical high-resolution STEM image. From this atomically resolved cross sectional view, the α-$In_2Se_3$ thin layer is found to be truly following the 3R stacking order with each unit cell contained three shifted QLs, which is identical with the schematic atom model as displayed in Figure 1a. In addition, we performed the Raman scattering measurements to show the sample uniformity and quality at macroscopic scale. Phonon modes centered at 90 $cm^{-1}$, 104 $cm^{-1}$, 182 $cm^{-1}$, 201 $cm^{-1}$ and 251 $cm^{-1}$ were observed in the Raman spectrum of typical samples (Supplementary Figure S1). The red shift of $A_1$ vibration mode to 104 $cm^{-1}$ indicates the phonon softening in the ferroelectric phase. All these phonon modes are in good agreement with the results from previous studies on ferroelectric 3R α-$In_2Se_3$.[21, 22, 26, 29] The Raman features, together with the AC-STEM results, exclusively confirm the 3R crystal structure nature of α-$In_2Se_3$ thin layers studied in this work.

As the first step, we verified the the ferroelectric polarization coupling in α-In$_2$Se$_3$ via the piezo force microscopy (PFM). The α-In$_2$Se$_3$ nanoflakes were post transferred onto conductive Au/SiO$_2$ substrate (details can be found in Methods). As shown in Supplementary Figure S3, the IP and OOP spontaneous ferroelectric domains are clearly visualized in the PFM phase images of α-In$_2$Se$_3$ few layers. The phase contrast between adjacent domains is observed to be 180°, indicating the antiparallel ordering of the electric dipoles in both the two directions. By comparing the IP and OOP PFM phase images, we found almost identical distribution of the ferroelectric domains. This strong correlation can be regarded as one evidence of the electric dipole inter-locking in α-In$_2$Se$_3$. However, when the polar axis of the studied ferroelectric is not strictly parallel or perpendicular to the PFM tip, same domain patterns will be produced in the IP and OOP directions as a result of the electric dipole projection. Therefore, to exclude such mechanism, we tested the simultaneous switching of the electric polarization in ferroelectric α-In$_2$Se$_3$ by an OOP electric field. Figure 1d to 1f present a square artificial ferrielectric domain established by +7 V vertical bias on one 14 nm thick sample. The surface topography measured after the writing process indicates that there is no damage to the sample under this bias. The flip of the OOP electric polarization is intuitively indicated by the strong phase contrast in the electrically written area as shown in Figure 1e. At the same time, we observed corresponding IP phase change in the same region (Figure 1f), indicating the concurrent switch of IP electric dipoles. The results from the artifical ferroelectric domains exclusively confirm the inherent coupling effect of IP and OOP ferroelectric polarization in 2D α-In$_2$Se$_3$.

To provide a proof-of-concept demonstration of the IP electric field control over the OOP electric polarization, we performed in-situ PFM study on a planar device of α-In$_2$Se$_3$. Figure 2a shows the schematic of our two terminal device (optical image and the topography can be found in Supplementary Figure S4). Before applying the in-plane voltage, we balanced the upward and downward electric polarizations via artifical domain engineering, which could help to the better observation of the coupling effect in α-In$_2$Se$_3$. Here, we define the OOP polarization degree of the device with $P = \frac{D_\downarrow - D_\uparrow}{D_\downarrow + D_\uparrow}$, where $D_\downarrow$ and $D_\uparrow$ acount the apperances of downward and upward electric polarization in the PFM measurements respectively. As shown in Figure 2b, a square downward artifical domain (bright area in the PFM phase image) is created with positive vertical PFM tip bias. Together with the spontaneous ferroelectric domains, the conducting channel of the device is divided into two oppositely polarized regions with almost equal size. Quantitatively, according to the statistics of $D_\downarrow$ and $D_\uparrow$ in Figure 2c, the P of the device at this initial state is negligible with the value of 5.2%. We then applied $\pm 40$ V IP pulsed voltage, equivalent to an electric field at $4\times10^4$ V cm$^{-1}$ in this device, to flip the OOP electric polarizations. After applying +40 V IP bias pulse, the area of downward ferroelectric domain was suppressed (Figure 2d), sugguesting the successful establishment of the upward electric polarization in the OOP direction. We found an increased P to 61.4% from the sharp contrast of $D_\downarrow$ and $D_\uparrow$ in Figure 2e. It should be highlighted that by -40 V IP voltage polarizing the device the domain pattern could be nearly restored to its fresh state with reduced P at 10.3% (Figure 2f & 2g). The reversible tuning on P via the planar bias validates the capability of the IP electric field control of the OOP electric polarization in α-In$_2$Se$_3$. Besides, we also observed clear OOP domain wall motion during the flipping process (Supplementary Figure S5), which was further evidence for this novle ferroelectricity control. It is worth nothing that, according to our previous PFM results, the swithcing electric field for the OOP ferroelectric polarization is at about $2\times10^5$ V cm$^{-1}$,[22] which is much larger than the case in the IP direction.

Before constructing the CPFET, we first quantitfy the gating ability of the OOP electric polarization in α-In$_2$Se$_3$. As studied earlier, due to the built-in electric field stemming from the electric polarization, large surface potential difference will be produced on the two sides of a given QL.[26] Therefore, when integrated with other vdW material forming vertical heterojunction or device, the polarity of the ferroelectric layer determines its electronic structure and transport property through the electrostatic doping or the modulation on the interfacial Schottky barrier.[22, 23] However, it is challenge to measure such difference directly from the top and bottom surfaces of ultra-thin vdW material in reality. Herein, as an indirect method, we perform study on opposite ferroelectric domains of α-In$_2$Se$_3$ with scanning Kelvin probe force microscopy (KPFM), which is senstive to the distributions of surface electric potential and charges. Figure 3a sketches the band alignments with respect to the vacuum level for the top side of α-In$_2$Se$_3$ at its two degenerate polarized states. As visualized in Figure 3b and 3c, in a typical sample with spontanous domains, we observed one to one correspondance of the surface potential mapping to the ferroelectric domain pattern. This correlation excludes the origin of the fluctuation in KPFM image from trapped charges or defects in α-In$_2$Se$_3$. Within each single domain, the contact potential signal is uniform. Between domains with antiparallel electric polarizations, the contact potential difference (CPD) were found to be 0.094 eV. At room temperature, under thermal equilibrium condition, the charge desnity within the same material follows $\Delta n = \exp(\frac{\Delta CPD}{k_b T})$, where $\Delta n$ represents the carrier density ratio, $k_b$ is the Boltzmann constant, and $T$ is the temperature. Therefore, with $\Delta CPD = 0.094\ eV$, the relative change of surface charge density between upward and downward electrically polarized α-In$_2$Se$_3$ is as high as 10 times, which is able to induce significant electrostatic doping in vertical heterostructures.

The CP-FET was constructed via stacking 2H-MoS$_2$ (bottom) and α-In$_2$Se$_3$ (top) vertically into heterojunction. In the device, MoS$_2$ and α-In$_2$Se$_3$ formed a cross bar like structure and were used as conducting channel and gate dielectric respectively. Figure 4a and 4b show the three-dimensional schematic view and optical image of our CP-FET (device #1). The as-prepared device is with four co-plannar eletrodes, among which two are used to read and the others are in charge of gating (or writting). The I-V characteristics of the CP-FET was measured by applying a relatively low voltage on the drain-source channel. The channel current ($I_{MoS_2}$) was linearly dependent on the bias (Supplementary Figure S6), indicating the Ohmic contacts between the metal electrode and MoS$_2$ few layers. To test the in-plane voltage gating effect in our CP-FET, we applied an transient IP voltage (V$_{in}$) on α-In$_2$Se$_3$ to control its OOP polarization which eventually induced the electrostatic doping in MoS$_2$, resulting the modulation on channel conductance. It should be noted that the MoS$_2$ channel was in open circuit when applying the IP polarizing vlotage on α-In$_2$Se$_3$. This helped to exclude the establishment of OOP polarization from the vertical bias generated on α-In$_2$Se$_3$ and MoS$_2$. When measuring the $I_{MoS_2}$, similar operation was applied on the gating electrodes. In the transfer characteristic measurement of the CP-FET, as a transient gating, the IP polarization process on α-In$_2$Se$_3$ was kept for 10 seconds before measuring the $I_{MoS_2}$. After 100 s for the electric polarization to get stable, a constant bias was applied on MoS$_2$ to detect the channel current $I_{MoS_2}$. Figure 4c shows the $I_{MoS_2}$ as the function of the IP gate voltage V$_{in}$. We observed clear hysteresis loop with conductance modulation as observed in previous 2D FeFETs.[13] As the V$_{in}$ was switched from -39 V to +39 V back and forth, the $I_{MoS_2}$ could be effectively tuned from 0.26 μA to 0.52 μA and vice versa with 200% change. This conductance variation is the result of the OOP electric polarization flipping by the IP electric field, which stems from the inherent coupling of the IP and OOP polarizations in α-In$_2$Se$_3$.

To study the dynamical evolution of OOP electric polarizations under the control of the IP voltage, we measured the time dependent channel current $I_{MoS_2}$ immediately after polarizing the α-In$_2$Se$_3$ with different V$_{in}$ pulses as shown in Figure 4d and 4e. With negative IP voltage polarization, the conductance gradually increased to a stable state, while the $I_{MoS_2}$ decreased after the initialization by positive IP voltage. Such different behaviors are well expected from the ferroelectric nature of α-In$_2$Se$_3$. For ferroelectrics, after the establishment of polarization, the ordering of electric dipoles will be disturbed by environmental thermal fluctuations and slowly decay to a stable state, resulting the so-called remnant electric polarization. Therefore, in the CP-FET, after the IP gate voltage polarizing, the conductance started to decay but with opposite directions for negative and positive bias. With a time span of 100 seconds, the device got to the constant conducting state. From the time dependent study, the CP-FET was found to be retentive, showing the non-volatile functionality. We summarized more detailed time evolution of the conductance as well as the electrical hysteresis in Supplementary Figure S7 and Supplementary Figure S9.

For the further check on the validity of our CP-FET, we carried out two control experiments. First of all, since all the electrical measurements were carried out in vacuum, the possibility that surface charge traps induce the above current modulation and the electrical hysteresis loop can be excluded. The hysterestic conduction modulation in the device was attributed to the screening effect of the electric polarizations of ferroelectrics. Therefore, artificially tuning the distance betweeen the ferroelectric layer and the semiconductor can significantly affect the generation of screening charges. Here, we fabricated a gapped CP-FET (termed as device #2) with $MoS_2$/hBN/α-$In_2Se_3$ vertical heterostructure as shown in Supplementary Figure S8. Ultra-thin hBN was used as buffer layer which increases the distance between the ferroelectric α-$In_2Se_3$ and $MoS_2$. Even in this type of gapped CP-FET, we observed the same electrical hysteresis but with relatively weak conductance modulation. It was well expected as a result of the imperfect screening of electric polarizations. Moreover, we made a device of $MoS_2$/$MoS_2$ homostructure to rule out the possible origin from electrostatically trapped charge or charge transfer between vdW layers (Supplementary Figure S11). With the same measuring sequence used in the study of CP-FET, the channel current in the bottom $MoS_2$ was found to be keeping constant when scanning the in-plane bias applied on the top $MoS_2$ layer. No electrical hysteresis was observed in this type of device.

## 3. Conclusion

In summary, we have presented a new approach to generate the electric field-effect with layered ferroelectric α-$In_2Se_3$ under room temperature. An in-plane voltage gated CP-FET with retentive switching effect have been demonstrated. This novel gating effect originates from the unique inter-locking of electric dipoles in α-$In_2Se_3$. It provides an unprecedented platform to develop novel functional nanoelectronic devices with vdW materials.

## 4. Experimental Section

*Sample preparation and topography measurements.* Bulk single crystal graphite, 2H-$MoS_2$, hBN, and α-$In_2Se_3$ used in this study were purchased from 2D Semiconductors

Inc. and Alfa Aesar Inc. respectively. Ultrathin vdW materials were prepared by the mechanical exfoliation onto Si substrate with 300 nm SiO$_2$ on top or transparent PDMS substrate (Gel-Pak, WF-60-X4) for stacking heterostructure. The topography and thickness of these ultra-thin samples were characterized by atomic force microscopy (AFM) (AIST-SmartSPM) at tapping mode.

*Raman spectroscopy characterization.* Raman spectra were taken with Horiba micro-Raman system (Labram HR Evolution). The wavelength of the laser excitation was 633 nm with on-sample power at around 100 μW. For α-In$_2$Se$_3$, as shown in Supplementary Figure S1, five characteristic Raman modes were observed to confirm its ferroelectric phase. For MoS$_2$ few layers, the $E_{2g}^1$ and A$_{1g}$ phonon modes were used to identify their layer number.[30] In the CP-FET (device #1), we found that the peak locations of $E_{2g}^1$ and A$_{1g}$ Raman modes of ultra-thin MoS$_2$ were centered at 383.8 cm$^{-1}$ and 405.2 cm$^{-1}$ respectivly with the wavenumber difference of 21.4 cm$^{-1}$ (Supplementary Figure S6), which confirmed the MoS$_2$ used in this device was bilayer.

*PFM measurements.* PFM measurements were performed with the same AIST-SmartSPM by using gold coated conducting Si tip under ambient condition. The conductive substrates used in this study were prepared through sputtering 20 nm thick gold film onto Si substrate with 300 nm SiO$_2$ on top.

*The CP-FET device fabrication and transport measurement.* MoS$_2$ few layers were transferred onto SiO$_2$/Si$^{++}$ by micromechanical exfoliation. We exfoliated a suitable few layer α-In$_2$Se$_3$ from bulk material. This few layer α-In$_2$Se$_3$ was then accurately transferred on to the top of MoS$_2$ bilayer via the all-dry transfer technique.[31] The cross bar like electrodes were fabricated by standard UV lithography and the following electron beam evaporation of 10 nm Ti and 70 nm Al. Keysight B2900 source meter was used to measure the I-V and transfer characteristics of our devices. All the measurements were performed in a vacuum chamber under the pressure at 1×10$^{-2}$ Pa.

## Acknowledgments

This work was supported by the National Key Research and Development Program of China (Grant No.2017YFA0205004, 2018YFA0306600, and 2017YFA0204904), the National Natural Science Foundation of China (Grant No.11674295, 11674299, 11374273, 11634011, and 51732010), the Fundamental Research Funds for the Central Universities (Grant No. WK2340000082, WK2030020032 and WK2060190084), Anhui Initiative in Quantum Information Technologies (Grant No. AHY170000), the Strategic Priority Research Program of Chinese Academy of Sciences (Grant No. XDB30000000), and the China Government Youth 1000-Plan Talent Program. This work was partially carried out at the USTC Center for Micro and Nanoscale Research and Fabrication.

## Author Contributions

H.Z. and W.Z. conceived the idea and supervised the research. Y. L., X. M., W. L., and H.L. prepared the samples. A.N, J.X., and Z.L. contributed to the high-resolution atomic structure identification with AC-STEM. Y.L., C.C. and W.L. fabricated the devices, carried out the transport, PFM, and Raman spectrum measurements of the CP-FETs. Y.L., C.C., and H.Z. analyzed the data, wrote the paper, and all authors commented on the manuscript.

## Author Information

The authors declare no competing financial interests. Correspondence and requests for materials should be addressed to Wenguang Zhu (wgzhu@ustc.edu.cn) and Hualing Zeng (hlzeng@ustc.edu.cn).

# Data availability

All relevant data are available from the authors.

Figures

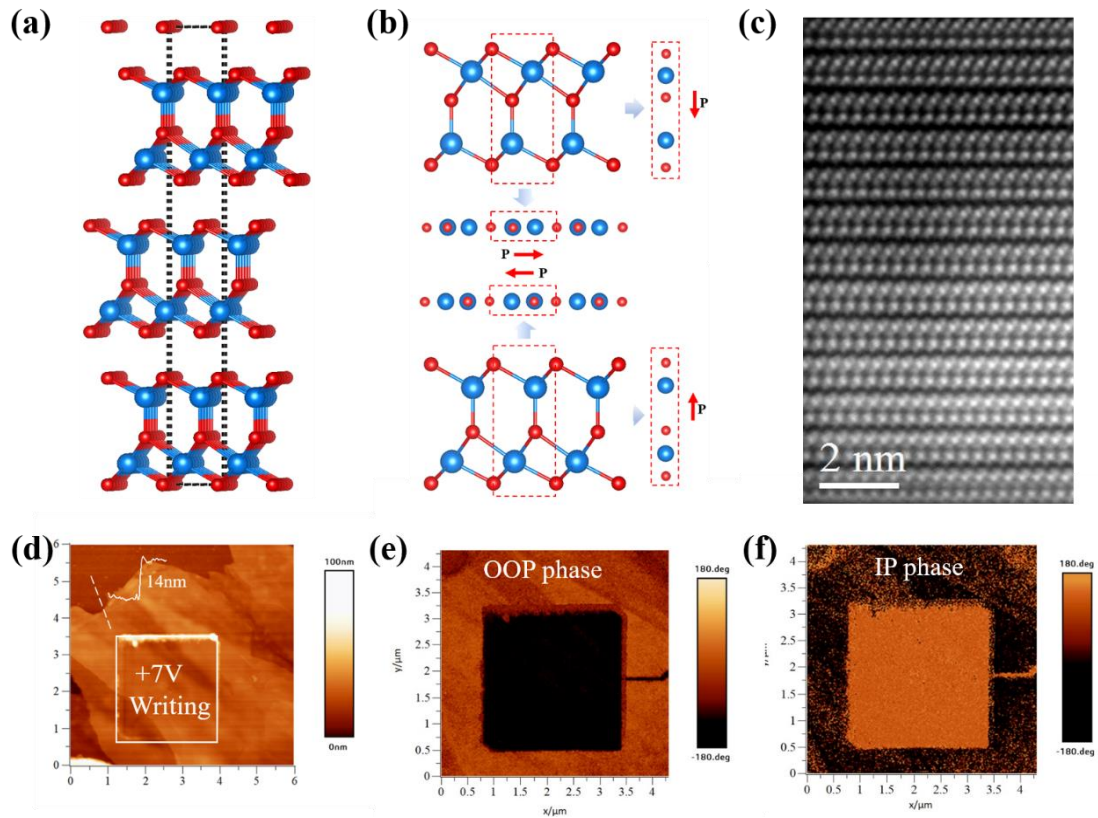

**Figure 1. Ferroelectric polarization locking in α-In₂Se₃.** a) Crystal structure of 3R α-In₂Se₃. Indium atoms and Selenium atoms are presented in blue and red respectively. b) Side views of the two oppositely polarized states of one QL α-In₂Se₃. Projection of the atoms into c and a (or b) axis indicated the spontaneously intercorrelated electric polarizations in α-In₂Se₃. The directions of the electric polarization were indicated by red arrows. c) Cross-sectional high-resolution STEM image of an α-In₂Se₃ nanoflake. Scale bar: 2 nm. d) Topography and e) & f) PFM measurements of 14 nm thick α-In₂Se₃. The ferroelectric domains in the OOP and IP directions are correlated.

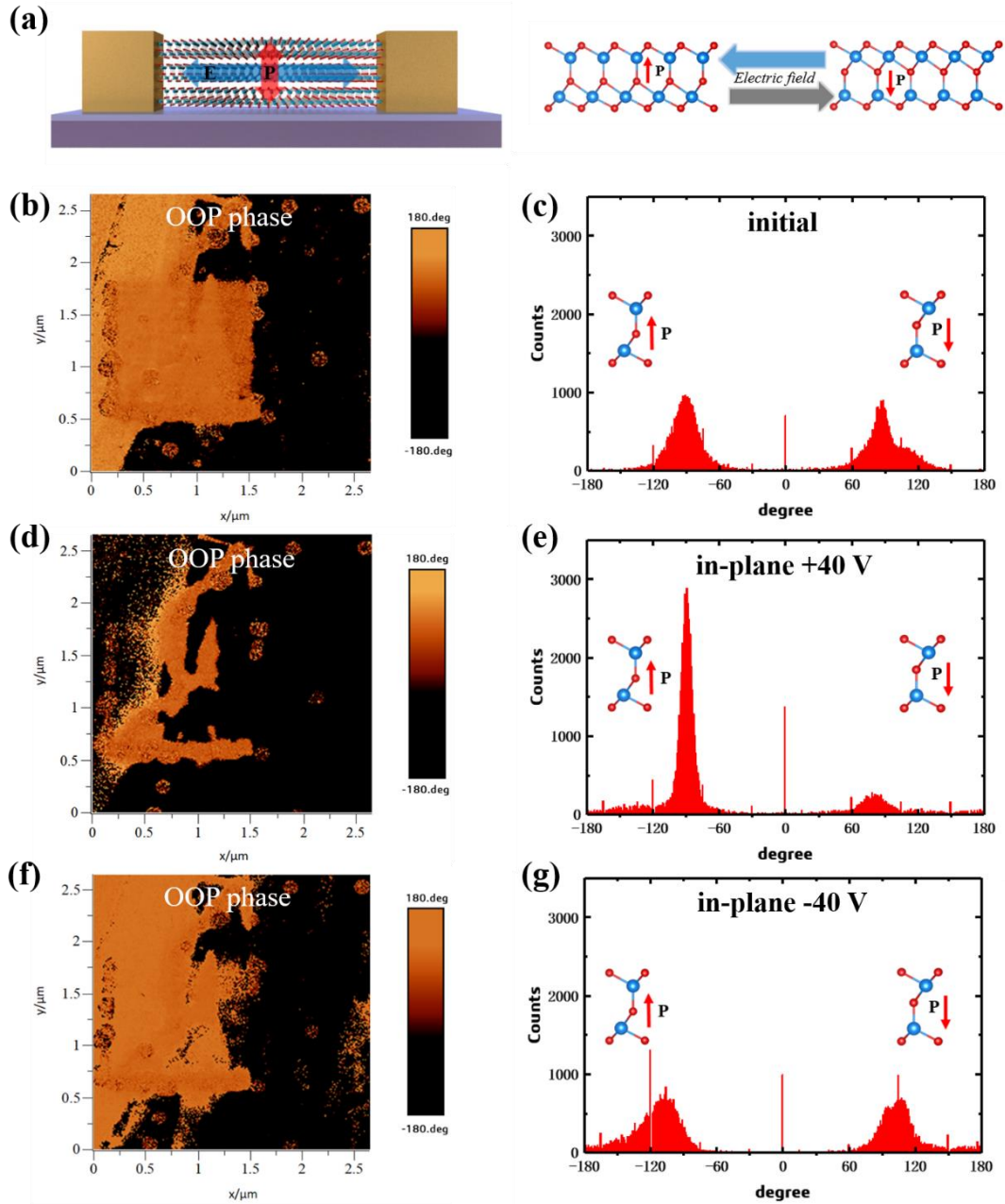

**Figure 2. In-plane electric field control of the out-of-plane electric polarization.** a) Schematic of the switching of OOP electric polarizations with IP electric field in a two-terminal device of ultra-thin α-In$_2$Se$_3$. b) – g). In-situ OOP electric polarization study with PFM. The OOP electric polarizations can be reversibly flipped by the IP bias of the proposed device in (a). The appearances of upward and downward electric polarizations in (b), (d), and (f) are counted in (c), (e), and (g) respectively.

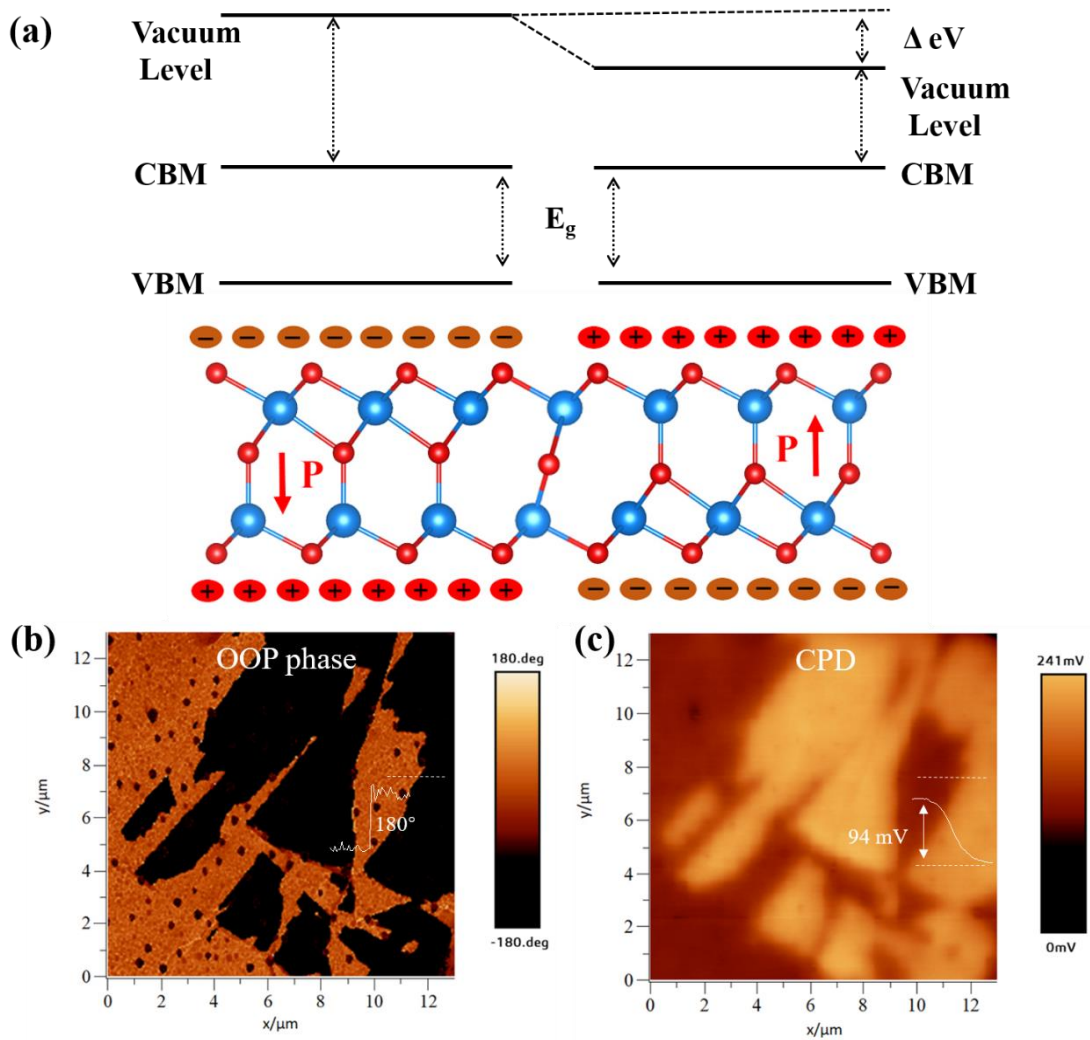

**Figure 3. Surface potential of ferroelectric domains.** a) The band alignments relative to vacuum level for downward and upward polarized α-$In_2Se_3$. The built-in electric field arising from the electric polarization leads to dramatic difference in surface potential. The positive and negative charges on the two sides of a QL stand for the polarization charges. b) Spontaneous OOP ferroelectric domains visualized by PFM. The phase profile of different ferroelectric domains is sketched by the dash line. A phase contrast of 180° is observed, which indicates the antiparallel directions of OOP polarization between the adjacent domains. c) The corresponding KPFM measurement on the ferroelectric domains in (b). The surface potential difference of antiparallel polarized OOP domains is found to be 94 mV.

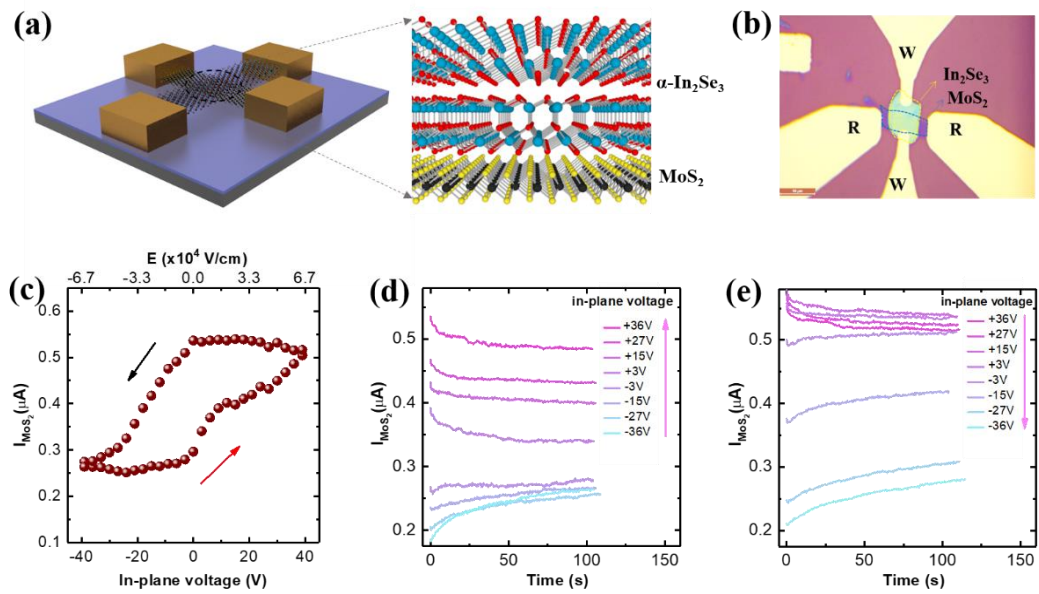

**Figure 4. Demonstration of the in-plane voltage gated field-effect transistor.** a) 3D schematic diagram of the CP-FET. The CP-FET is fabricated by vertically stacking MoS$_2$ and α-In$_2$Se$_3$ thin layers. The zoomed area shows the crystal structure of the MoS$_2$/α-In$_2$Se$_3$ heterostructure. b) Optical image of the CP-FET device. c) The hysteretic in-plane ferroelectric gating of the CP-FET. The conductance of the MoS$_2$ channel is tuned by the IP gate voltage with clear electrical hysteresis loop. d) & e) Time dependent conductance study of the CP-FET. The variation of the channel current follows the decay of the electric polarizations. The arrow indicates the measuring sequence.